# Synthesis, Characterization and Ageing of $MgB_2$

A. Serquis[1], R. Schulze[2], Y. T. Zhu[1], J.Y. Coulter[1], D. E. Peterson[1], N.O. Moreno[3], P. G. Pagliuso[3], S.S. Indrakanti[4], V. F. Nesterenko[4] and F. M. Mueller[1]
[1] Superconductivity Technology Center, MS K763
[2] Materials Technology Metallurgy Group, MS G755
[3] Condensed Matter and Thermal Physics, MS K764
Materials Science and Technology Division, Los Alamos National Laboratory, Los Alamos, NM 87544, USA
[4] Department of Mechanical and Aerospace Engineering, University of California, San Diego, La Jolla, CA 92093

## ABSTRACT

We studied the influence of sample preparation and defects in the superconducting properties samples using atomic ratios of Mg:B=1:1 and Mg:B=1:2. Samples were characterized by SEM, and XRD, and the magnetization properties were examined in a SQUID magnetometer. The presence of Mg vacancies was determined by Rietveld analysis. Most of the samples exhibited sharp superconducting transitions with Tcs between 37- 39 K.

We found a strong correlation between the crystal strain and the Tc. This strain was related to the presence of Mg vacancies. In addition, results showed that some samples degraded with time when exposed to ambient conditions. In these samples the Tc did not change with time, but the superconducting transition became broader and the Meissner fraction decreased. This effect was only present in samples with poor connectivity between grains and smaller grain sizes. The degradation was related to a surface decomposition as observed by X-ray Photoelectron Spectroscopy. No correlation was found between this effect and the presence of Mg vacancies.

## INTRODUCTION

Since the discovery of superconductivity in $MgB_2$ at 39 K [1], the highest Tc observed for a non-copper-oxide bulk superconductor, considerable progress has been made in the understanding of the fundamental properties of this material. $MgB_2$ appears to be a suitable candidate for superconducting technologies currently based on Nb-alloys wires and films. However, to make practical devices using $MgB_2$, it is essential to have a stable material. Taking into account that most high-temperature superconductors are highly sensitive to water and moist air it is important to explore the influence of exposure $MgB_2$ to air and water. Zhai et al.[2] have studied the degradation of superconducting properties in $MgB_2$ films by exposure to water and have observed that the Tc(onset) of the films remains unchanged throughout the degradation process. Nevertheless, there are no detailed studies on the chemical stability of $MgB_2$, specially regarding its sensitivity to $O_2$, $H_2O$ and $CO_2$ in ambient conditions.

Besides, different sample preparation methods lead to materials with different superconducting properties [3-5]. It is interesting to note that the origin of lower Tc of the in-situ prepared films, the possible non-stoichiometry and the effect of possible O contamination remain topics of debate. Cooper et al [6] proposed that the compounds with the $AlB_2$-type structure could be non-stoichiometric. While Rogado et al [7] found a small dependency between a possible Mg deficiency and the Tc, Bordet et al [4] suggested that this system should be extremely sensitive to Mg non-stoichiometry. Zhao et al [8] reported that the superconducting temperature of $MgB_2$ changes significantly with different starting compositions of Mg:B, which was possibly caused by Mg vacancies.

We study the influence of sample preparation and defects (i.e. Mg vacancies or the presence of strain-inducing MgO inside the $MgB_2$ grains) in the superconducting properties of $MgB_2$ samples. We found that most of the samples exhibit sharp superconducting transitions but Tc varied between 37- 39 K in magnetization measurements. The results also show that some samples can undergo to degradation with time when exposed to ambient conditions.

These results will be discussed in terms of differences in the structural features (i.e. grain sizes) and crystalline defects such as Mg vacancies.

## EXPERIMENTAL

Samples A-D were prepared using an atomic ratio of Mg:B = 1:1, varying starting materials, and/or synthesis temperatures and times, under flowing Ar or inside a quartz tube (see Table I). Details of sample preparation are reported elsewhere [9-10]. For comparison, we have prepared another sample of $MgB_2$ (Sample E), starting with a stoichiometric mixture of Boron and Magnesium following the standard procedure of other authors as described elsewhere [10].

Powder X-ray diffraction data were collected using a Scintag XDS2000 θ–θ powder diffractometer. All samples were nearly single phase with small quantities of MgO. The Mg vacancies and strain were determined by Rietveld analysis of powder X-ray data using the computer program GSAS [11] as a mixture of $MgB_2$ (space group P6/mmm), MgO (Fm-3m) and Si (Fd3m) standard.

A superconducting quantum interference device (SQUID) magnetometer (Quantum Design) was used to measure the susceptibility of the samples at 10 Oe. The direct current (dc) resistivity as a function of temperature was measured through the standard 4-probes method in a computer-controlled data logger system. No transport measurements could be made in samples D and E, because the first was not well sintered and the second was in powder form.

A sample for X-ray Photoelectron Spectroscopy (XPS) was prepared by hot isostatic pressure (HIP), since this technique is advantageous in producing fully dense bulk $MgB_2$ samples suitable for surface analysis. All high resolution XPS spectra were collected with a Physical Electronics 5600ci XPS system equipped with an Omni IV hemispherical capacitor analyzer in retarding mode of operation and a pass energy of 23.5 eV; details of sample preparation and collection data are given elsewhere [6]. Atomic concentrations were calculated for the XPS spectra using the following relative sensitivity factors determined for the XPS system used: B1s 0.171, C1s 0.314, O1s 0.733, Mg2p 0.167.

**Table I -** Synthesis parameters for $MgB_2$ samples.

| Sample | Starting Materials & Processing States | Mg:B | Temp | Time | Cooling | Annealing |
|---|---|---|---|---|---|---|
| A | Mg turnings + B powder pellets | 1:1 | 900°C | 2 h | slow | Ar flux |
| B | Mg + B powder mixture pellets | 1:1 | 900°C | 2 h | slow | Ar flux |
| C | Mg + B powder mixture pellets | 1:1 | 750°C | 7 h | slow | Ar flux |
| D | Mg turnings B + powder pellets | 1:1 | 900°C | 1 h | quenched | Quartz tube |
| E | Mg turnings + B powder | 1:2 | 950°C | 2 h | fast | Quartz tube |

## RESULTS

In Fig. 1(a) and (b) are shown the magnetization and resistivity data of the as-made samples. It can be observed that Samples A, B and C have very sharp transitions but the Tc onset change from ~39 K to ~37 K. The broadening in the transition corresponding to samples D and E are related to a poor connectivity and a small grain size, as it was confirmed by SEM observations.

To determine the origin of the variations in the Tc, we analyzed the defects present in these samples. From the X-ray measurements we observed that samples with lower Tc present a broadening in the diffraction peaks.(see Fig. 2). It is well known [12-13] that the x-ray peak broadening can be caused by both crystallite size and lattice strain. Strain broadening can be caused by any lattice defects, such as vacancies, interstitials, substitutions and stacking faults. In this analysis, a pseudo-Voight function was used to fit the Bragg diffraction peaks [9]. To refine the Mg occupancy in the $MgB_2$ phase, we first determined its approximate value by manually checking the goodness of fit, $\chi^2$, as a function of Mg occupancy. Table II summarizes the structural features of each sample.

It is clear from data in Table II and Fig. 2 that there is a strong correlation between the Tc and the lattice strain. As it was explained with more detail in ref.[9], Mg deficiency is likely to be the cause of the observed strain.

To study the effect of the chemical stability when samples are exposed to ambient conditions, we have measured the same samples after several periods of time. In Fig 3 (a), (b) and (c) are plotted the ZFC dc magnetization vs. T after preparation and after an aging time for the $MgB_2$ samples D, and E, respectively. We have not included samples A, B and C, which was prepared with excess Mg and well sintered with larger grain sizes, as they remain stable during several months under ambient conditions. In contrast, samples D and E undergo a slow degradation, the Tc does not change but the superconducting fraction is reduced and the superconducting transition becomes broader.

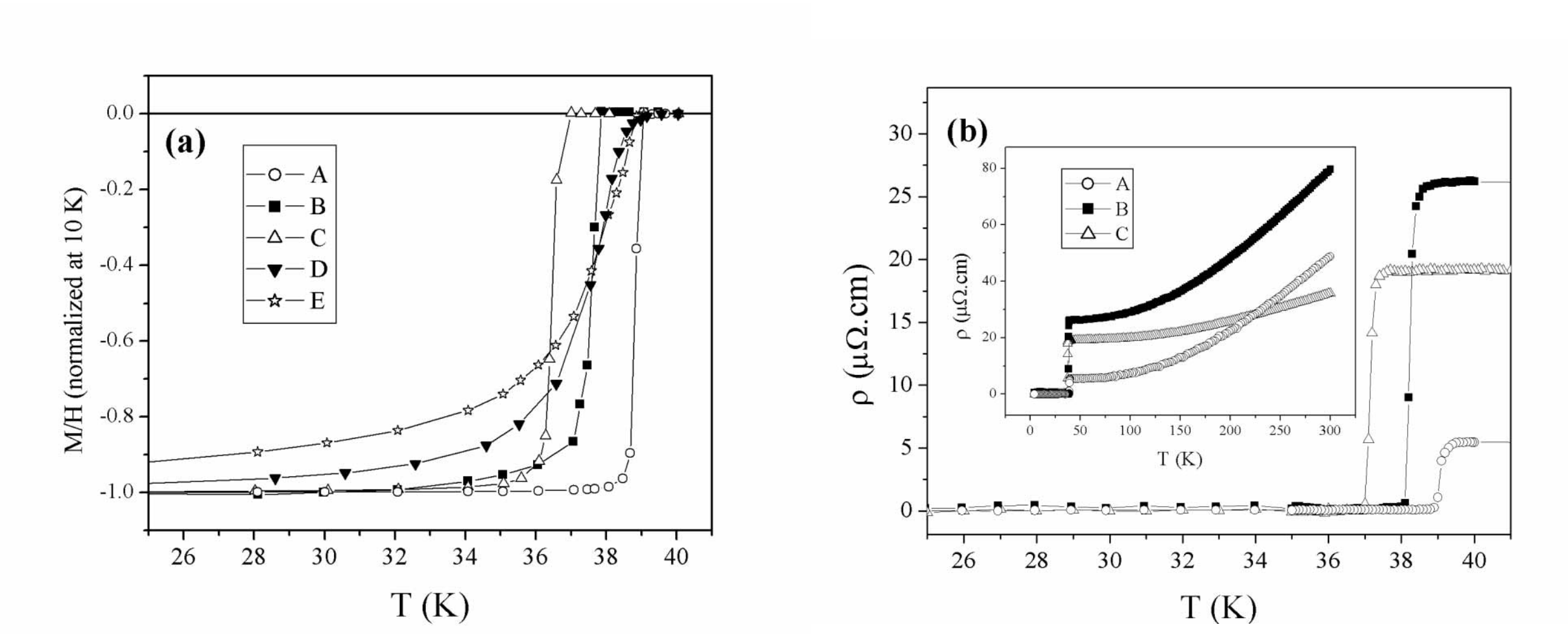

**Fig. 1(a)** Magnetic susceptibility vs. temperature of all $MgB_2$ as-made samples. A constant magnetic field of 10 Oe was applied. **(b)** dc resistivity vs. temperature for samples A, B and C.

**Table II –** Structural features of each sample. Tc corresponds to the magnetization data.

| Sample | Tc | Δ Tc (10-90%) | Grain size (SEM) | Strain | Mg Occupancy |
|---|---|---|---|---|---|
| A | 38.9 ± 0.2 | 0.4 K | 0.5-5.0 μm | 0.309(5) | 0.994(5) |
| B | 37.8 ± 0.3 | 0.8 K | 0.5-1.0 μm | 0.665(6) | 0.974(6) |
| C | 36.6 ± 0.3 | 0.6 K | 0.3-1.0 μm | 1.079(8) | 0.946(7) |
| D | 38.8 ± 0.2 | 4.3 K | 0.1-0.5 μm | 0.407(7) | 0.983(7) |
| E | 38.9 ± 0.2 | 10.3 K | 0.2-1.0 μm | 0.233(6) | 0.993(5) |

Figure 4 (a) shows a composite of the spectra acquired for the Mg KLL line for the air aging experiment. One might consider the sputter cleaned $MgB_2$ spectrum to be the intrinsic Mg KLL spectrum from $MgB_2$, although the intensity at about 306 eV probably contains some "oxidized" contribution from the starting impurity level of O in the sample. With increasing air exposure, the low energy feature at 302.1 eV, which is a signature of the intrinsic $MgB_2$, drops in intensity, while the feature at 306.6 eV, which we believe to be a signature of oxidized Mg, increases in intensity. The analysis of the XPS spectrum of $MgB_2$ as a function of air exposure time in comparison with standards of $Mg(OH)_2$, $MgCO_3$, and MgO suggests that the major impurity in the $MgB_2$ from air exposure is a Mg hydroxide, $Mg(OH)_x$, of indeterminate stoichiometry [10]. It is likely, however, that the oxidized magnesium is of mixed chemical environment, having some components of MgO and Mg oxycarbon in addition to the Mg hydroxide.

The increase of C and O content while B atomic concentration is reduced in the surface of $MgB_2$ samples is clearly seen in Fig. 4(b). This suggests that the B in the $MgB_2$ matrix is being replaced by O and C containing species through reaction with $O_2$, $CO_2$ (and possibly other C containing species), and $H_2O$ in the air exposure.

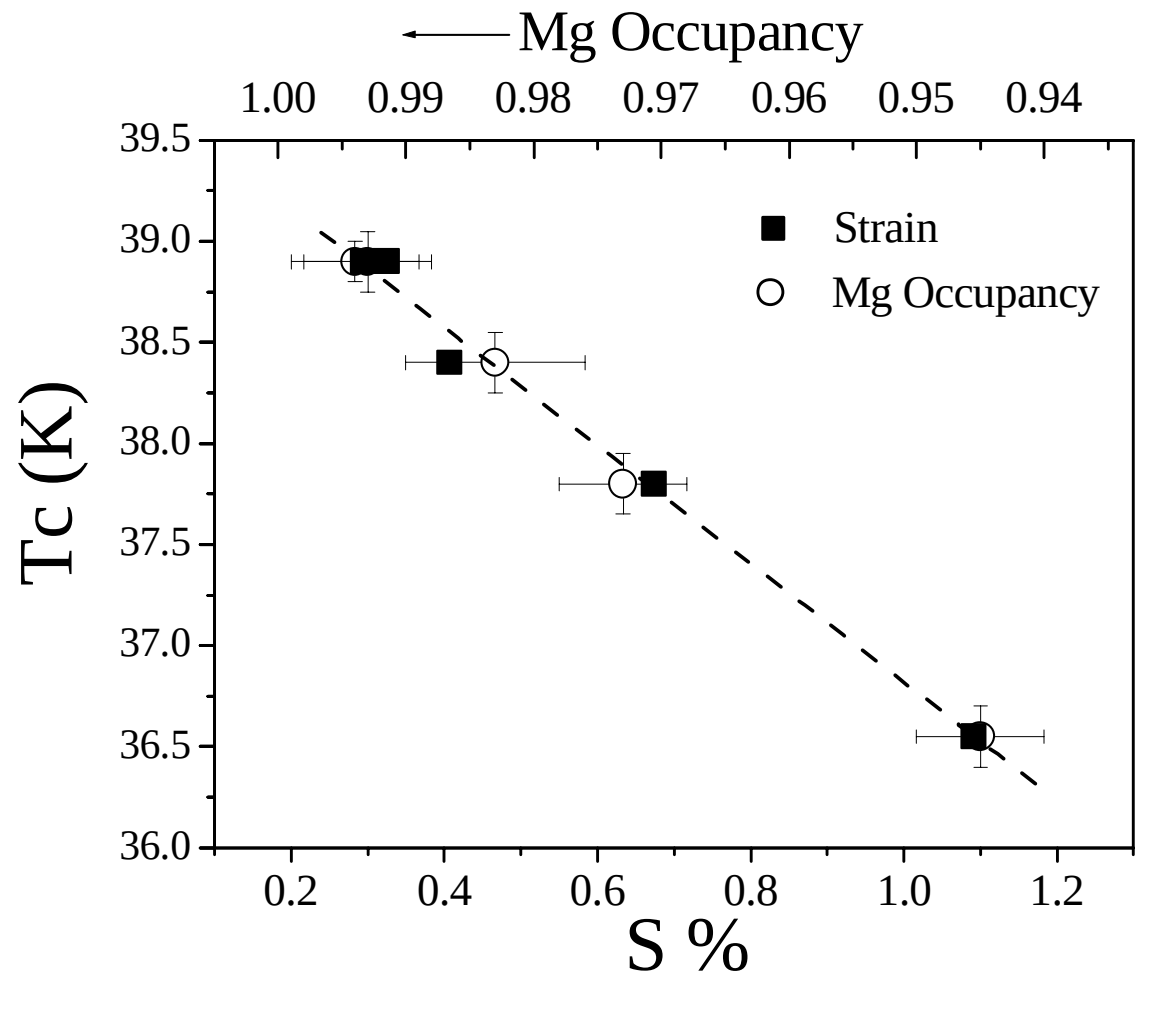

**Fig. 2-** The derived isotropic strain and the Mg occupancy as a function of Tc.

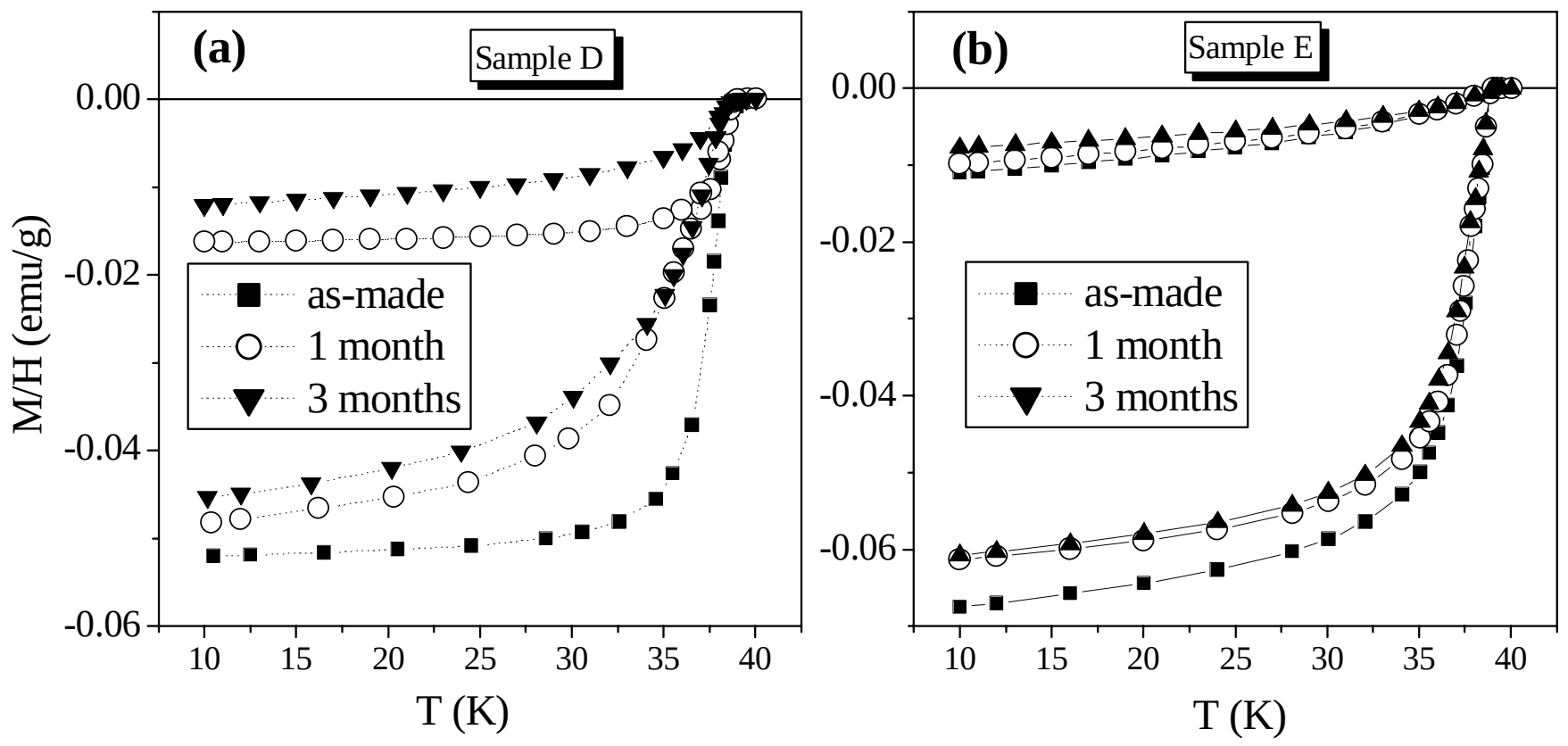

**Fig. 3-** Magnetic susceptibility of MgB2 samples as a function of temperature for different ageing times. (a) Sample D and (b) Sample E.

The sample used for the surface analysis has the same behavior as Sample A, i.e., remaining stable for several months. Taking into account that the hydroxylated surface layer analyzed in the XPS experiment is ~20 Å thick (based on the ratio of the $MgK_{LL}$ line intensity contribution from the intrinsic $MgB_2$ and the contribution from the Mg hydroxide intensity in the 1278 hour air exposed sample), the volume affected in a grain with a radius of 5 μm is only 0.1%, while for a radius of 1-0.1 μm is 0.6-6 %, respectively. Therefore, samples with grain sizes larger than 5 μm or well sintered, in such a way that the grain surface is not in contact with air, will not be affected by these surface reactions. Schmidt et al.[14] suggested that chemical modifications of the surface (e.g. $MgB_2$ reacts with water) could potentially produce a layer with a lower Tc. These authors also pointed out that if Mg vacancies are present in the near-surface regions, $H^+$ could enter the compound interstitially or as a replacement for missing $Mg^{2+}$ in the $MgB_2$ structure or it could be formed MgO[14]. As we can see in Table II, Samples D an E have Mg occupancies higher than 0.99 and we did not find any degradation in samples B and C with Mg vacancies up to 5 % [9]. However, it is necessary to take into account that if the Mg vacancies are present only in the surface, the XRD analysis cannot detect them, because it is a bulk technique. The Mg excess in the initial composition of the samples can contribute to avoid the presence of Mg vacancies, but also plays a role in the formation of well-sintered samples.

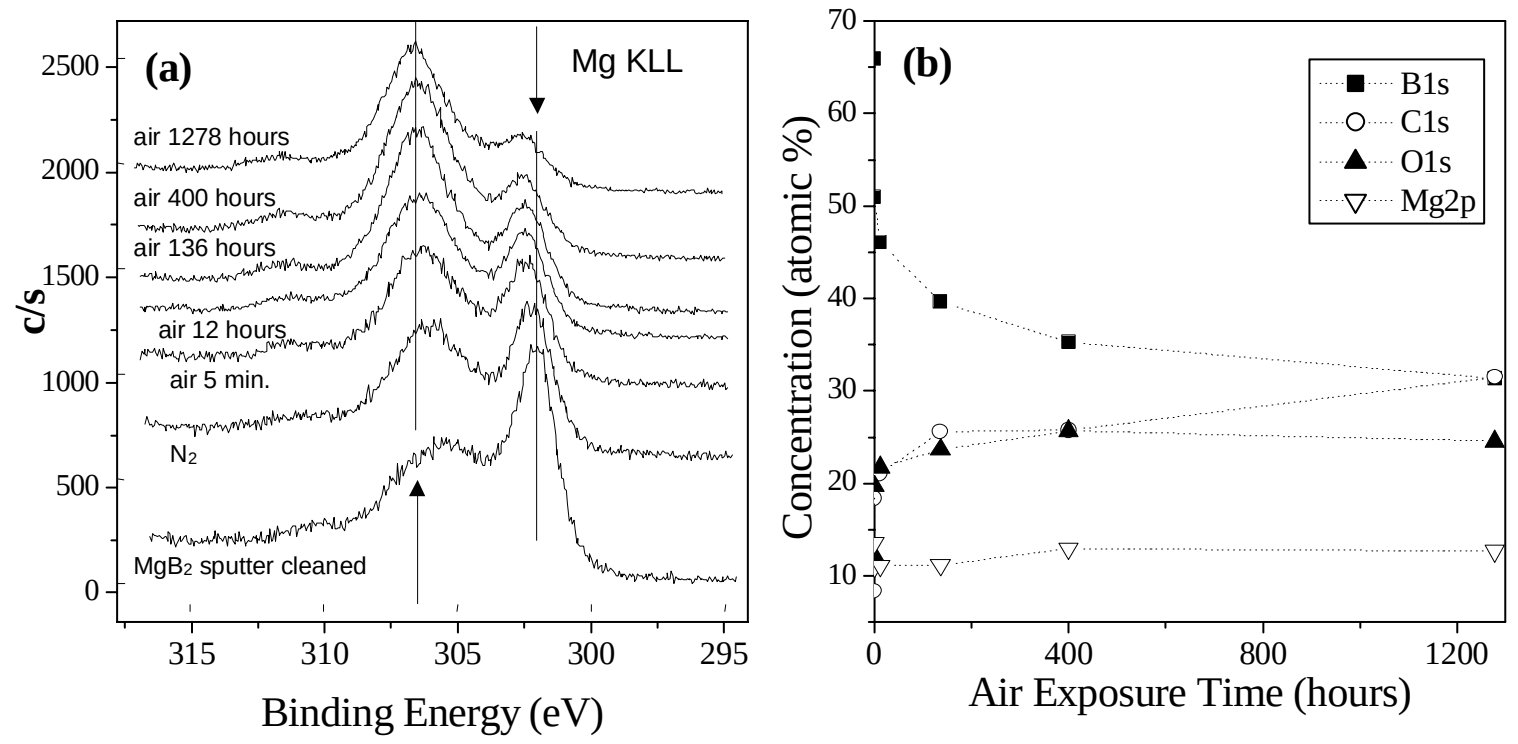

**Fig. 4 (a)** XPS spectrum of MgKLL for different air exposure times. **(b)** Relative atomic concentration (%) as a function of air exposure time.

## CONCLUSIONS

We have studied the influence of sample preparation and structural features in the superconducting properties and chemical stability of $MgB_2$ samples to obtain materials that may be suitable for applications. We have found a strong correlation between the crystal strain and the Tc. This strain is related with the presence of Mg vacancies. The presence of Mg excess in the synthesis in an open atmosphere under Ar flux contributes in the formation of well-sintered samples. Our results showed that some samples can degrade with time when exposed to ambient conditions. In these samples, the Tc did not change with time, but the superconducting transition became broader and the Meissner fraction decreases. This effect is only present in samples with poor connectivity between grains and smaller grain sizes. The degradation is related to a surface decomposition as observed by X-ray Photoelectron Spectroscopy. No correlation was found between this effect and the presence of Mg vacancies. The upshot is that it is possible to make $MgB_2$ samples, which do not degrade with time.

## ACKNOWLEDGMENT

This work was performed under the auspices of the US DOE Office of Energy Efficiency and Renewable Energy, as part of its Superconductivity for Electric Systems Program.